\documentclass{llncs}
\usepackage{graphicx}
\usepackage{amsmath}
\usepackage{amsfonts}
\usepackage{amssymb}
\newtheorem{algorithm}{Algorithm}
\newenvironment{proof}[1][Proof]{\textbf{#1.} }{\ \rule{0.5em}{0.5em}}

\begin{document}
\title{Algorithms for Boolean Function Query Properties}
\author{Scott Aaronson\thanks{Supported by an NSF Graduate\ Fellowship. \ Work done at Bell Laboratories / Lucent Technologies.} }
\institute{Computer Science Division, UC Berkeley \\ Berkeley, CA USA 94720-1776 \\ \texttt{aaronson@cs.berkeley.edu}
}
\date{}
\maketitle
\begin{abstract}
We present new algorithms to compute fundamental properties of a Boolean function given in truth-table form. \
Specif\hspace{0pt}ically, we give an $O(N^{2.322}\log N)$\ algorithm for block sensitivity,
an $O(N^{1.585}\log N)$ algorithm\ for
`tree decomposition,' and an $O(N)$ algorithm for `quasisymmetry.' \ These algorithms are based on new insights 
into the structure of Boolean functions that may be of independent interest.
\ We also give a subexponential-time algorithm for the space-bounded quantum
query complexity of a Boolean function. \ To prove this algorithm correct, we develop a theory of 
limited-precision representation of unitary operators,
building on work of Bernstein and Vazirani.

\end{abstract}
\section{Introduction}
\vspace{-0.4cm} \noindent The query complexity of Boolean
functions, also called black-box or decision-tree complexity, has been well studied for years
\cite{bbcmw,bw,nisan,ns,rubinstein,sw}. \ 
Numerous Boolean function properties relevant to query complexity have
been def\hspace{0pt}ined, such as sensitivity, block sensitivity, randomized and quantum query complexity, and degree as a real 
polynomial. \ But many open questions remain
concerning the relationships between the properties. \ For example, are
sensitivity and block sensitivity polynomially related? \ How small can
quantum query complexity be, relative to randomized query
complexity? \
Lacking answers to these questions, we may wish to gain insight into them
by using computer analysis of small Boolean functions. \ But to perform
such analysis, we need ef\hspace{0pt}f\hspace{0pt}icient algorithms to compute the properties in
question. \ Such algorithms are the subject of the present paper.

Let $f:\left\{  0,1\right\}  ^{n}\rightarrow\left\{  0,1\right\}  $ be a
Boolean function, and\ let $N=2^{n}$ be the size of the truth table of $f$.
\ We seek algorithms that have modest running
time as a function of $N$, given the truth table as input. \ The following
table lists some properties important for query complexity, together with the
complexities of the most ef\hspace{0pt}f\hspace{0pt}icient algorithms for them of which we know.  In the table, `LP' stands for linear programming reduction.

\bigskip
\noindent\begin{tabular}
[c]{l @{\hspace{0.5cm}} l @{\hspace{0.5cm}} l}
\textbf{Query Property} & \textbf{Complexity} & \textbf{Source} \\
Deterministic query complexity $D(f)$ & $O(N^{1.585}\log N)$ & \cite{glr} \\
Certif\hspace{0pt}icate complexity $C(f)$ & $O(N^{1.585}\log N)$ & \cite{czort} \\
Degree as a real polynomial $\operatorname*{deg}(f)$ & $O(N^{1.585}\log N)$ & This paper \\
Approximate degree $\widetilde{\operatorname*{deg}}(f)$ & About $O(N^{5})$ & Obvious (LP) \\
Randomized query complexity $R_{0}(f)$ & About $O(N^{7.925})$ & Obvious (LP) \\
Block sensitivity $\operatorname*{bs}(f)$ & $O(N^{2.322}\log N)$ & This paper \\
Quasisymmetry & $O(N)$ & This paper \\
Tree decomposition & $O(N^{1.585}\log N)$ & This paper \\
Quantum query complexity $Q_{2}(f)$ & Exponential & Obvious \\
$Q_{2}(f)$ with $O(\log n)$-qubit restriction & $O(N^{\operatorname*{polylog}(N)})$ & This paper%
\end{tabular}

\bigskip

There is also a complexity-theory rationale for studying algorithmic problems
such as those considered in this paper. \ 
Much effort has been devoted to f\hspace{0pt}inding Boolean function properties that do not naturalize in the sense of
Razborov and Rudich \cite{rr}, and that might therefore be useful for proving circuit lower bounds.  \ In our view,
it would help this effort to have a better general understanding of the complexity of problems on Boolean function truth tables---both
upper and lower bounds. \ Such problems have been considered since the 1950's \cite{trakhtenbrot}, but basic open questions remain, especially in the setting of circuit complexity \cite{kc}.  \ This paper addresses the much simpler setting of query complexity.

We do not know of a polynomial-time algorithm to f\hspace{0pt}ind quantum query
complexity; we raise this as an open problem. \ However, even f\hspace{0pt}inding
quantum query complexity via exhaustive search is nontrivial, since it involves representing unitary operators
with limited-precision arithmetic.
\ The problem is deeper than that of approximating unitary
gates with bounded error, which was solved by Bernstein and Vazirani
\cite{bv}. \ In Section 7 we resolve the problem, and
give an $O(N^{\operatorname*{polylog}(N)})$ constant-factor approximation
algorithm for bounded-error quantum query complexity if the memory of the
quantum computer is restricted to $O(\log n)$ qubits.

We have implemented some of the algorithms discussed in this paper in a
linkable C library \cite{aaronson}, which
is available for download.
\vspace{-0.3cm} \section{Preliminaries\label{prelimsec}}
\vspace{-0.4cm} \noindent A Boolean function $f$\ is a total function from $\left\{  0,1\right\}
^{n} $\ onto $\left\{  0,1\right\}  $. \ We use $V_{f}$ to denote the set of variables of $f$, and use $X$,
or alternatively $x_{1},\ldots,x_{n}$, to denote an input to $f$. \  If $X$ is an input, $\left|X\right|$ denotes the Hamming weight
of $X$; if $S$ is a set, $\left|S\right|$ denotes the cardinality of $S$.
Particular Boolean functions to which we ref\hspace{0pt}er are $\operatorname*{AND}_{n}$, $\operatorname*{OR}_{n}$, and $\operatorname*{XOR}_{n}$,
the $\operatorname*{AND}$, $\operatorname*{OR}$, and $\operatorname*{XOR}$ functions respectively on $n$ inputs.
\vspace{-0.3cm} \section{Previous Work\label{basicprops}}
\vspace{-0.4cm} \noindent To our knowledge, no algorithms for block sensitivity, quasisymmetry, tree decomposition,
or quantum query complexity have been previously published. \ But algorithms 
for simpler query properties have appeared in the literature.

Given a Boolean function $f$, the {\it deterministic query complexity} $D(f)$ is the minimum height of a decision tree 
representing $f$. \
Guijarro et al. \cite{glr} give a simple $O(N^{1.585}\log N)$ dynamic programming algorithm to compute $D(f)$. 
\ That $f$ is given as a truth table is crucial: if $f$ is non-total and only the inputs for which $f$ is def\hspace{0pt}ined are given,
then computing $D(f)$ is (when phrased as a decision problem) \textsf{NP}-complete \cite{hr}.

The {\it certif\hspace{0pt}icate complexity} $C(f)$ is the maximum, over all inputs $X$, of the minimum number of input bits
needed to prove the value of $f(X)$. 
\ Equivalently, $C(f)$ is the minimum height of a nondeterministic decision tree for $f$.
\ Czort \cite{czort} gives an $O(N^{1.585}\log N)$\ algorithm to compute $C(f)$. \
Again, if $f$ is not given as a full truth table, then computing $C(f)$ is \textsf{NP}-complete \cite{hjlt}.

Let $\deg(f)$\ be the minimum degree of an $n$-variate real multilinear
polynomial $\rho$ such that, for all $X\in\left\{  0,1\right\}  ^{n}$,
$\rho(X)=f(X)$. \ 
The following
lemma, adapted from Lemma 4 of \cite{bw}, is easily seen to yield an $O(N^{1.585}\log N)$ dynamic programming
algorithm for $\deg(f)$. \ Say that a function
obeys the {\it parity property} if the number of inputs $X$ with odd parity for
which $f(X)=1$ equals the number of inputs $X$ with even parity for which
$f(X)=1$.

\begin{lemma}[Shi and Yao]
\label{paritylemma}$\deg(f)$\ equals the size of the
largest restriction of $f$ for which the parity property fails.
\end{lemma}
\vspace{-0.3cm} \section{\label{bssec}Block Sensitivity}
\vspace{-0.4cm} \noindent Block sensitivity, introduced in \cite{nisan}, is a Boolean function
property that is used to establish lower bounds.  There are several open problems that an ef\hspace{0pt}f\hspace{0pt}icient
algorithm for block sensitivity might help to investigate
\cite{nisan,bbcmw,bw}.

Let $X$ be an input to Boolean function $f$, and let $B$ (a {\it block}) be a
nonempty subset of $V_{f}$.  Let $X(B)$ be the input obtained from $X$ by f\hspace{0pt}lipping the bits of $B$.

\begin{definition}
A block $B$ is {\normalfont sensitive} on $X$ if $f(X)\neq f(X(B))$, and {\normalfont minimal} 
on $X$ if $B$ is sensitive and no proper sub-block $S$ of $B$ is sensitive.
Then the block sensitivity $\operatorname*{bs}_{X}(f)$
of $X$ is the maximum number of disjoint minimal (or equivalently, sensitive) blocks on $X$.
\ F\hspace{0pt}inally $\operatorname*{bs}(f)$ is the maximum of
$\operatorname*{bs}_{X}(f)$\ over all $X$.
\end{definition}

The obvious algorithm to compute $\operatorname*{bs}(f)$ (compute $\operatorname*{bs}_{X}(f)$ for each $X$ using
dynamic programming, then take the maximum) uses $\Theta(N^{2.585}\log N)$ time.
 \ Here we show how to reduce the complexity to $O(N^{2.322}\log N)$ by exploiting the structure of minimal blocks.
\ Our algorithm has two main stages: one to identify minimal blocks and store them for fast lookup, another to compute
$\operatorname*{bs}_{X}(f)$ 
for each $X$ using only minimal blocks. \ The analysis proceeds by showing that no Boolean function has too many
minimal blocks, and therefore that if the algorithm is slow for some inputs (because of an abundance of minimal blocks), then it
must be faster for other inputs.

\begin{algorithm}
\textbf{(computes $\operatorname*{bs}(f)$)}
\label{bs}For each input $X$:
\begin{enumerate}
\item  Identify all sensitive blocks of $X$; place them in an AVL tree
$T$.
\item  Loop over all sensitive blocks $B$ in $T$\ in lexicographic order 
($\left\{  x_{1}\right\}  ,\left\{  x_{2}\right\}  ,$ $\left\{  x_{1}%
,x_{2}\right\}  ,\left\{  x_{3}\right\}  ,\ldots$). \  For each block $B$, loop
over all $2^{n-|B|}-1$ possible blocks $C$ that properly contain
$B$.\ \ Remove from $T$ all such blocks $C$ that are in the tree; such blocks have
been identif\hspace{0pt}ied as non-minimal.
\item  Create $2^{n}-1$ lists, one list $L_{S}$ for each nonempty subset $S
$ of variables. \ Then, for each minimal block $B$ in $T$, insert a copy of
$B$ into each list $L_{S}$ such that $B\subseteq S$. \ The result is that,
for each $S$, $L_{S}=2^{S}\cap T$, where $2^{S}$ is the power set of $S$.
\item  Let a {\normalfont state} be a partition $\left(  P,Q\right)  $ of $V_{f}$. \ The set $P$
represents a union of disjoint minimal blocks that have already been selected;
the set $Q$ represents the set of variables not yet selected. \ Then
$\operatorname*{bs}_{X}(f)=\theta\left(  \emptyset,V_{f}\right)$, where
$\theta\left(  P,Q\right)  $ is def\hspace{0pt}ined via the recursion
$\theta\left(  P,Q\right) \overset{\triangle}{=}1+\max_{B\in L_{Q}%
}\theta\left(  P\cup B,Q-B\right).  $
Here the maximum evaluates to $0$ if $L_{Q}$ is empty. \ Compute
$\theta\left(  P,Q\right)  $ using depth-f\hspace{0pt}irst recursion,  caching
the values of $\theta\left(  P,Q\right)  $ so that each
needs to be computed only once.
\end{enumerate}
\end{algorithm}

\noindent The block sensitivity is then the maximum of $\operatorname*{bs}_{X}(f)$ over
all $X$.

Let $m(X,k)$ be the number of minimal blocks of $X$ of size $k$. \ The
analysis of Algorithm \ref{bs}'s running time depends on the following lemma, which
shows that large minimal blocks are rare in any Boolean function.

\begin{lemma}
\label{bslemma}$
{\sum_{X}}
m(X,k)\leq2^{n-k+1}\tbinom{n}{k}$.
\end{lemma}
\begin{proof}
\nobreak The number of positions that can be occupied by a minimal block of size $k$ is
$\tbinom{n}{k}$ for each input, or $2^{n}\tbinom{n}{k}$ for all inputs.
\ Consider an input $X$ with a minimal block $B=\left\{b_{1},\ldots,b_{k}\right\}$ of size $k>1$. \
Block $B$ has $2^{k}-1$ nonempty subsets; label them $S_{1},\ldots,S_{2^{k}-1}$. \
By the minimality of $B$, for each $S_{i}$ the input $X(S_{i})$ has 
$\left\{b_{1}\right\},\ldots,\left\{b_{k}\right\}$ as minimal blocks if $S_{i}=B$,
and $B-S_{i}$ as a minimal block if $S_{i}\neq B$. \ Therefore $X(S_{i})$ cannot have $B$ 
as a minimal block. \ So of the $2^{n}\tbinom{n}{k}$ positions, only one out of 
$2^{k}$ can be occupied by a minimal block of size $k$.  When $k=1$ an additional factor of $2$ is needed, since $X(B)$ has $B$ as a minimal block.
\end{proof}

\begin{theorem}
Algorithm \ref{bs} takes $O(N^{2.322}\log N)$ time.
\end{theorem}
\begin{proof}
Step 1 takes time $O(N^{2}\log N)$, totaled over all inputs.
Let us analyze step 2, which identif\hspace{0pt}ies the minimal blocks. \ For each input
$X$, every block $B$ that is selected is minimal, since each non-minimal block
in $A$ was removed in a previous iteration. \ Furthermore, for each block $B$
the number of removals of $C$ blocks is less than $2^{n-|B|}$.
\ Therefore the total number of removals is at most
 \[ {\textstyle\sum_{X}} {\textstyle\sum_{k=0}^{n}} m(X,k)2^{n-k}
  = {\textstyle\sum_{k=0}^{n}} \left[  2^{n-k} {\textstyle\sum_{X}} m(X,k)\right] 
  \leq {\textstyle\sum_{k=0}^{n}} 2^{2n-2k+1}\tbinom{n}{k} \]
which sums to $2N^{\log_{2}5}$.  Since each removal takes $O\left(  \log N\right)  $\ time, the total
time is $O\left(  N^{2.322}\log N\right)  $.

We next analyze step 3, which creates the $2^{n}-1$\ lists $L_{S}$. \ Since
each minimal block $B$ is contained in $2^{n-|B|}$\ sets of variables, the
total number of insertions is at most 
$  {\sum_{k=0}^{n}} m(X,k)2^{n-k} $
for input $X$.  So the time is $O\left(  N^{2.322}\log
N\right)  $\ by the previous calculation.

F\hspace{0pt}inally we analyze step 4, which computes block sensitivity using the minimal
blocks. \ Each $\theta\left(  P,Q\right)  $\ evaluation is
performed at most once, and involves looping through a list of minimal blocks
contained in $Q$, with each iteration taking $O(\log N)$\ time. \ For each block $B$, the number of distinct $(P,Q)$ 
pairs such that $B\subseteq
Q$\ is at most $2^{n-|B|}$. \ Therefore, again, the time for each input $X$ is at most 
$ (\log N){\sum_{k=0}^{n}} m(X,k)2^{n-k} $
and a bound of $O\left(N^{2.322} \log N\right)  $\ follows.
\end{proof}
\vspace{-0.3cm} \section{\label{quasisymsec}Quasisymmetry}
\vspace{-0.4cm} \noindent A Boolean function $f(X)$ is {\it symmetric} if its output depends only on $\left|
X\right|  $. \ Query complexity is well understood for
symmetric functions: for example, for all non-constant symmetric $f$, the deterministic query complexity is $n$ and
the zero-error quantum query complexity is $\Theta\left(  n\right)$ \cite{bbcmw}. \ Thus, a program for analyzing Boolean functions might f\hspace{0pt}irst check whether a
function is symmetric, and if it is, dispense with many
expensive tests. 
We call $f$ {\it quasisymmetric} if some subset of input bits can be negated to make $f$ symmetric. \ For example, $f=\operatorname*{OR}\left( x_{1}, \urcorner x_{2} \right)$ is quasisymmetric but not symmetric.  \ There is an obvious
$O(N^{2})$ algorithm to test quasisymmetry; here we sketch a linear-time algorithm.

Call a restriction of $f$ a $z$-\textit{left-restriction}
if each variable $v_{i}$\ is f\hspace{0pt}ixed if and only if $i\leq z$. \ Our algorithm
recurses through all $z$-left-restrictions: when it is called on restriction $R$,
it calls itself recursively on $R_{x_{z+1}=0}$\ and
$R_{x_{z+1}=1}$. \ If either of these is not quasisymmetric, then the
algorithm returns failure; otherwise, the algorithm tries to f\hspace{0pt}it the restrictions
together in such a way that $R$ itself is seen to be quasisymmetric. \ It does this by testing whether
$R_{x_{z+1}=0}\left(\left| X \right|\right) = R_{x_{z+1}=1}\left(\left| X \right| \pm 1 \right)$, with separate
routines for the special cases in which $R_{x_{z+1}=0}$ or $R_{x_{z+1}=1}$ is a constant function or a $\operatorname*{XOR}$ or $1-\operatorname*{XOR}$ function. \ If the
f\hspace{0pt}itting-together process succeeds, then the algorithm returns both the output
of $R$ (encoded in compact form, as a symmetric function)\ and the set of
input bits that must be f\hspace{0pt}lipped to make $R$ symmetric. \ Crucially, these
return values occupy only $O(n-z)$\ bits of space. \ 
The algorithm also has subroutines to handle the special cases in which $R$ is a $\operatorname*{XOR}$ function 
or a constant function. \
In these cases $R$ is symmetric no matter which set of input bits is f\hspace{0pt}lipped. 
Since the time used by each invocation is linear in $n-z$, the total time used is 
$ {\sum_{z=0}^{n}} 2^{z}(n-z)=O(N). $
The following lemma shows that the algorithm\ deals with all of the ways in
which a function can be quasisymmetric, which is key to the
algorithm's correctness.

\begin{lemma}
Let $f$ be a Boolean function on $n$ inputs. \ If two distinct (and non-complementary) sets
of input bits $A$ and $B$ can be f\hspace{0pt}lipped to make $f$ symmetric, then $f$ is either
$\operatorname*{XOR}_{n}$, $1-\operatorname*{XOR}_{n}$, or a constant function.
\end{lemma}
\begin{proof}
\nobreak Assume without loss of generality that $B$ is empty. \ Then $A$ has cardinality less than $n$. \ We know that
$f(X)$\ depends only on $\left|  X\right|  $, and also that it depends only on
$ \left/  X\right/  \overset{\triangle}{=} {\sum_{i=1}^{n}} \kappa(x_{i}) $
where $\kappa(x)=1-x$ if $x_{i}\in A$\ and $\kappa(x)=x$\ otherwise. \ Choose any
Hamming weight $0\leq w\leq n-2$,\ and consider an input $Y$ with $\left|Y\right|=w$ and with two variables 
$v_{i}$ and $v_{j}$ such that $v_{i}\in A$, $v_{j}\notin A$, and $Y(i)=Y(j)=0$. \ Let $Z=Y_{Y(i)=1,Y(j)=1}$.
\ We have $\left|  Z\right|  =\left|  Y\right|  +2$, but on the other hand 
$\left/Z\right/  =\left/  Y\right/$, so $f(Y)=f(Z)$ by symmetry.
\ Again applying symmetry, $f(P)=f(Q)$ whenever $\left|  P\right|=w$ and $\left|  Q\right|  =w+2$. \ Therefore $f$ is either
$\operatorname*{XOR}_{n}$, $1-\operatorname*{XOR}_{n}$, or a constant function.
\end{proof}
\vspace{-0.3cm} \section{\label{primesec}Tree Decomposition}
\vspace{-0.4cm} \noindent Many of the Boolean functions of most interest to query complexity are
naturally thought of as trees of smaller Boolean functions: for
example, AND-OR trees and majority trees. \ Thus, given a function $f$,
one of the most basic questions we might ask is whether it has a tree
decomposition and if so what it is. \ In this section we def\hspace{0pt}ine a sense in which
every Boolean function has a unique tree decomposition, and we prove its uniqueness. \ We also sketch an
$O(N^{1.585}\log N)$ algorithm for f\hspace{0pt}inding the decomposition.

\begin{definition}
A {\normalfont distinct variable tree} is a tree in which
\begin{enumerate}
\item [(i)]Every leaf vertex is labeled with a distinct variable.
\item[(ii)] Every non-leaf vertex $v$ is labeled with a Boolean function
having as many variables as $v$ has children, and depending on all of its variables.
\item[(iii)] Every non-leaf vertex has at least two children.
\end{enumerate}
\end{definition}

\noindent Such a tree represents a Boolean function in the obvious way. \ We call the tree
{\it trivial} if it contains exactly one vertex.

A {\it tree decomposition} of $f$ is a separation of $f$ into the smallest possible
components, with the exception of $\left(\urcorner\right)\operatorname*{AND}_{k}$,
$\left(\urcorner\right)\operatorname*{OR}_{k}$, and $\left(\urcorner\right)\operatorname*{XOR}_{k}$\ components
(where $\left(\urcorner\right)$ denotes possible negation), which are
left intact. \ The choice of $\operatorname*{AND}$, $\operatorname*{OR}$, and $\operatorname*{XOR}$ components is not arbitrary; 
these are precisely the three components that ``associate,'' so that, for example,
$\operatorname*{AND}\left(x_{1}, \operatorname*{AND}\left(x_{2},x_{3}\right)\right) = \operatorname*{AND}\left(\operatorname*{AND}\left(x_{1}, x_{2}\right), x_{3}\right)$. \ Formally:

\begin{definition}
A tree decomposition of $f$ is a distinct variable tree representing $f$
such that:
\begin{enumerate}
\item[(i)]No vertex is labeled with a function $f$ that can be represented by
a nontrivial tree, unless $f$ is $\left(  \urcorner\right)
\operatorname*{AND}_{k}$, $\left(\urcorner\right)  \operatorname*{OR}_{k}$,
or $\left(  \urcorner\right)  \operatorname*{XOR}_{k}$\ for some $k$.
\item[(ii)]No vertex labeled with $\left(\urcorner\right)\operatorname*{AND}_{k}$ has a child labeled with
$\operatorname*{AND}_{l}$.
\item[(iii)]No vertex labeled with $\left(\urcorner\right)\operatorname*{OR}_{k}$ has a child labeled with
$\operatorname*{OR}_{l}$.
\item[(iv)]No vertex labeled with $\left(\urcorner\right)\operatorname*{XOR}_{k}$ has a child labeled with
$\left(\urcorner\right)\operatorname*{XOR}_{l}$.
\item[(v)]Any vertex labeled with a function that is constant on all but one input is labeled with
$\left(\urcorner\right)\operatorname*{AND}_{k}$ or
$\left(\urcorner\right)\operatorname*{OR}_{k}$.
\end{enumerate}
\end{definition}

Let {\it double-negation} be the operation of negating the output of a
function at some non-root vertex $v$, then negating the corresponding input of
the function at $v$'s parent. \ Double-negation is a trivial way to obtain
distinct decompositions. 
This caveat aside, we can assert uniqueness:

\begin{theorem}
\label{unique}Every Boolean function has a unique tree decomposition, up to double-negation.
\end{theorem}
\begin{proof}[Proof]
Given a vertex $v$ of a distinct variable tree, let $L(v)$\ be the
set of variables in the subtree of which $v$ is the root. \ Assume that
$f$\ is represented by two distinct tree decompositions, $T$ and $Z$, such
that $T$ has a vertex $v_{T}$\ and $Z$ has a vertex $v_{Z}$\ with $L(v_{T})$
and $L(v_{Z})$ incomparable (i.e. they intersect, but
neither contains the other).\ \ Then let $A=L(v_{T})-L(v_{Z})$, $B=L(v_{Z}%
)-L(v_{T})$, $I=L(v_{T})\cap L(v_{Z})$, and $U=V_{f}-L(v_{T})-L(v_{Z})$.  The crucial lemma is the following.

\begin{lemma}
\label{tiz}$f$ is a function of $t\left(  A\right)  $, $i\left(  I\right)  $,
$z\left(  B\right)  $, and $U$, for some Boolean functions $t$, $i$, and $z$.
\end{lemma}

\begin{proof}
We can write $T$ as $T_{|U}\left[  t_{AI}\left(
A,I\right)  ,B\right]  $, where $t_{AI}$\ is Boolean;
similarly we can write $Z$ as $Z_{|U}\left[  A,z_{IB}\left(  I,B\right)
\right]  $. \ We have that, for all settings of $U$, 
$T_{|U}\left[  t_{AI}\left(  A,I\right)  ,B\right]  =Z_{|U}\left[
A,z_{IB}\left(  I,B\right)  \right]$.
\ Consider a restriction that f\hspace{0pt}ixes all the variables in $B$.
\ This yields 
$T_{|U,B}\left[  t_{AI}\left(  A,I\right)  \right]  =Z_{|U}\left[
A,z_{|B}\left(  I\right)\right].$ 
Therefore, for all restrictions of $B$, $t_{AI}$\ depends on only a
single bit obtained from $I$, namely $z_{|B}\left(  I\right)  $. \ So we can
write $t_{AI}\left(  A,I\right)  $\ as $t_{\alpha}\left(  A,z_{|B}\left(
I\right)  \right)  $ for some Boolean $t_{\alpha}$---or even more strongly as
$t_{\alpha}\left(  A,t_{\beta}\left(  I\right)  \right)  $, since we know that
$t_{AI}$ does not depend on $B$. \ By analogous reasoning we can write
$z_{IB}\left(  I,B\right)  $\ as $z_{\alpha}\left(  z_{\beta}\left(  I\right)
,B\right)  $ for some functions $z_{\alpha}$\ and $z_{\beta}$. \ So we have 
$
T_{|U}\left[  t_{\alpha}\left(  A,t_{\beta}\left(  I\right)  \right)
,B\right]  =Z_{|U}\left[  A,z_{\alpha}\left(  z_{\beta}\left(  I\right)
,B\right)  \right].
$ 
Next we restrict $A\cup B$, obtaining 
$
T_{|U,B}\left[  t_{\alpha|A}\left(  t_{\beta}\left(  I\right)  \right)
\right]  =Z_{|U,A}\left[  z_{\alpha|B}\left(  z_{\beta}\left(  I\right)
\right)\right],
$ 
which implies that, for some functions $T_{|U,B}^{\prime}$\ and $Z_{|U,A}%
^{\prime}$, 
$
T_{|U,B}^{\prime}\left[  t_{\beta}\left(  I\right)  \right]  =Z_{|U,A}%
^{\prime}\left[  z_{\beta}\left(  I\right)  \right]  .
$ 
This shows that $t_{\beta}\left(  I\right)  $ and $z_{\beta}\left(  I\right)
$\ are equivalent up to negation of output, since $T$ and $Z$ must depend on $I$ for some restriction of
$A\cup B$. \ So we have 
$
T_{|U}\left[  t_{i\left(  I\right)  }\left(  A\right)  ,B\right]
=Z_{|U}\left[  A,z_{i\left(  I\right)  }\left(  B\right)\right].
$
for some Boolean functions $i(I)$
(henceforth simply $i$), $t_{i}$, and $z_{i}$ ($i\in\left\{
0,1\right\}  $). \ Next we restrict $A$ and $i$: 
$
T_{|U,A,i}\left[  B\right]  =Z_{|U,A}\left[  z_{i|i}\left(  B\right)\right].
$
Thus, for all restrictions of $A$ and $i$, $T$ depends on only a single bit
obtained from $B$, which we'll call $\widehat{z}_{i}\left(  B\right)  $ (and
which can be taken equal to $z_{i|i}\left(  B\right)  $). \ Note that 
$\widehat{z}_{i}$\ does not depend on $A$. \ Analogously, for both
possible restrictions of $i$, $Z$ depends on only a single bit obtained from
$A$, which we'll call $\widehat{t}_{i}\left(  A\right)  $. \ So we can write 
$
T_{\ast|U}\left[  t_{i}\left(  A\right)  ,\widehat{z}_{i}\left(  B\right)
\right]  =Z_{\ast|U}\left[  \widehat{t}_{i}\left(  A\right)  ,z_{i}\left(
B\right)  \right]
$
where $T_{\ast|U}$ and $Z_{\ast|U}$\ are two-input Boolean functions. \ We
claim that $\widehat{z}_{0}=\widehat{z}_{1}$\ and $\widehat{t}_{0}=\widehat
{t}_{1}$.

There must exist a setting $u$ of
$U$ such that $T_{\ast|U=u}$ depends on both $t_{i}$ and $\widehat{z}_{i}$.
\ Suppose there exists a setting $b$ of $B$ such that $\widehat{z}_{0}\left(
b\right)  \neq\widehat{z}_{1}\left(  b\right)  $. \ $t_{i}$\ must be a
nonconstant function, so
f\hspace{0pt}ind a constant $c$ such that $T_{\ast|U=u}\left[
c,\widehat{z}_{i}\left(  B\right)  \right]$ depends on $\widehat{z}_{i}$,
and choose a setting for $A$ and $i$\ such that $t_{i}\left(  A\right)  =c$.
\ (If $T_{\ast|U=u}$\ is a $\operatorname*{XOR}$ function, then either
$c=0$ or $c=1$ will work, whereas if $T_{\ast|U=u}$\ is an
$\operatorname*{AND}$ or $\operatorname*{OR}$ function, then only one value of $c$ will work.) \ For
$T_{\ast|U=u}$\ to be well-def\hspace{0pt}ined, we need that whenever $t_{i}\left(
A\right)  =c$, the value of $i$ is determined (since $\widehat{z}_{i}$\ has no
access to $i$). \ This implies that $t_{i}$\ has the form $t\left(  A\right)
\wedge i$\ or $t\left(  A\right)  \wedge\urcorner i$\ for some function $t$.
\ Therefore $T_{\ast|U=u}$\ can be written as $T_{tiz|U=u}\left[  t\left(
A\right)  ,i,\widehat{z}_{i}\left(  B\right)  \right]  $ for some function
$T_{tiz|U=u}$.

Now repeat the argument for $Z_{\ast|U=u}$. \ We obtain that $Z_{\ast|U=u}%
$\ can be written as $Z_{tiz|U=u}\left[  \widehat{t}_{i}\left(  A\right)
,i,z\left(  B\right)  \right]  $\ for some functions $Z_{tiz|U=u}$.and $z$. \ Therefore 
$
T_{tiz|U=u}\left[  t\left(  A\right)  ,i,\widehat{z}_{i}\left(  B\right)
\right]  =Z_{tiz|U=u}\left[  \widehat{t}_{i}\left(  A\right)  ,i,z\left(
B\right)\right].
$
So we can take $\widehat{z}_{i}\left(  B\right)  =z\left(  B\right)  $
(equivalently $\widehat{t}_{i}\left(  A\right)  =t\left(  A\right)  $), and
write $T_{tiz|U=u}$\ (or $Z_{tiz|U=u}$) as 
\linebreak $
T_{tiz|U=u}\left[  t\left(  A\right)  ,i\left(  I\right)  ,z\left(  B\right)
\right].
$
\bigskip
\end{proof}

We now prove the main theorem: that $f$ has a unique tree decomposition, up to double-negation. \
From Lemma \ref{tiz}, $v_{T}$ effectively has as inputs the two bits $t(A)$ and $i(I)$, and $v_{Z}$ the
two bits $i(I)$ and $z(B)$.  Thus we can check, by enumeration,
that either
$v_{T}$\ and $v_{Z}$ are labeled with the same function, and that
function is either\ $\operatorname*{AND}_{k}$, $\urcorner\operatorname*{AND}%
_{k}$, $\operatorname*{OR}_{k}$, or $\urcorner\operatorname*{OR}_{k}$; or
$v_{T}$\ and $v_{Z}$\ are both labeled with either $\operatorname*{XOR}
_{k}$\ or $\urcorner\operatorname*{XOR}_{k}$. \
(Note that $k$ can be different for $v_{T}$\ and for $v_{Z}$.)

In either
case, for all $u$ there exists a function $T_{tiz|U=u}$, taking $t\left(
A\right)  $, $i\left(  I\right)  $, and $z\left(  B\right)  $\ as input, that
captures all that needs to be known about $A\cup I\cup B$. \ Furthermore,
since $t_{AI}$\ and $z_{IB}$\ do not depend on $u$, neither does $T_{tiz|U=u}$,
and we can write it as $T_{tiz}$. \ Let $v_{m}$\ be the unique
vertex in $T$ such that $L\left(  v_{m}\right)  $\ contains $A\cup I\cup
B$\ and is minimal among all $L\left(  v_{i}\right)  $\ sets that do so. \ If
$v_{m}$ is labeled with $\left(  \urcorner\right)  \operatorname*{AND}_{k}$,
$\left(  \urcorner\right)  \operatorname*{OR}_{k}$, or $\left(  \urcorner
\right)  \operatorname*{XOR}_{k}$, then $v_{T}$\ cannot be a vertex of $T$.
 \ If
$v_{m}$ is labeled with some other function, then $L\left(  v_{m}\right)  \neq
A\cup I\cup B$\ and the function at $v_{m}$\ is represented by a nontrivial
tree. \ 
Either way we obtain a contradiction.

Now that we have ruled out the possibility of incomparable subtrees, we can
establish uniqueness. \ Call a set $V\subseteq V_{f}$
\textit{unif\hspace{0pt}iable} if there exists a vertex $v_{\ast}$, in some
decomposition of $f$, such that $L\left(  v_{\ast}\right)  =V$. \ Let
$C$\ be the collection of all unif\hspace{0pt}iable sets. \ We have established that
no pair $V_{1}$, $V_{2} \in C$ is incomparable: either $V_{1}\cap V_{2}%
=\phi$, $V_{1}\subseteq V_{2}$, or $V_{2}\subseteq V_{1}$. \ We claim that any
decomposition must contain a vertex $v_{i}$\ with $L\left(
v_{i}\right)  =V_{i}$ for every $V_{i}\in C$. \ For suppose that
$V_{i}$\ is not represented in some decomposition $F$. \ Certainly
$V_{i}\neq V_{f}$, so let $V_{P}$\ be the \textit{parent set} of $V_{i}$\ in
$F$: that is, the unique minimal set such that $V_{i}\subset V_{P}$ and there
exists a vertex $v_{P}$\ in $F$ with $L\left(  v_{P}\right)  =V_{P}$. \ Then
the function at $v_{P}$\ is represented by a nontrivial tree,
containing a vertex $v_{i}$\ with $L\left(  v_{i}\right)  =V_{i}$---were it
not, then $v_{i}$\ could not be a vertex in any decomposition. \ 
Furthermore, the function at $v_{P}$\ cannot be $\left(
\urcorner\right)  \operatorname*{AND}_{k}$, $\left(  \urcorner\right)
\operatorname*{OR}_{k}$, or $\left(  \urcorner\right)  \operatorname*{XOR}%
_{k}$. \ If it were, then again $v_{i}$ could not be a vertex in any decomposition, 
since it would need to be labeled correspondingly with $\left(
\urcorner\right)  \operatorname*{AND}_{k}$, $\left(  \urcorner\right)
\operatorname*{OR}_{k}$, or $\left(  \urcorner\right)  \operatorname*{XOR}%
_{k}$. \
Having determined the unique set of vertices that comprise any tree decomposition, 
the vertices' labels are also determined up to double-negation.
\end{proof}

We now sketch an algorithm to construct the tree decomposition. \ 
In a distinct variable tree, let $L(v)$ be the set of variables in the subtree of which $v$ is the root.
\ Then given
a subset $G$ of $V_{f}$, we can clearly decide in linear time whether a
distinct variable tree representing $f$ could have a vertex $u$ with $L(u)=G$.
 \ So we can construct a decomposition in $O\left(  N^{2}\right)$ time,
by checking whether a
vertex $u$\ could have $L(u)=G$\ for each subset $G$ $\subseteq
V_{f}$\ \ satisfying $2\leq\left|  G\right|  \leq n-1$.

The key insight for reducing the time to $O(N^{1.585}\log
N)$ is to represent each restriction by a concise
\textit{codeword}, which takes up only $O\left(  n\right)  $ bits rather than
$2^{\left|  G\right|  }$ bits.\ \ We create the codewords recursively,
starting with the smallest restrictions and working up to larger ones. \ The
codewords need to satisfy the following conditions:
\begin{itemize}
\item Two restrictions $g$ and $h$ over the same set of variables get mapped to identical codewords if and only if $g=h$.
\item If $g$ is the negation of $h$, then this fact is easy to tell given the codewords of $g$ and $h$.
\item If a restriction is constant, then this fact is also easy to tell given its codeword. 
\end{itemize}
We can satisfy this condition by building up a binary tree of restrictions
at each recursive call, then assigning each restriction a codeword based on
its position in the tree. \ For all $G\neq\emptyset$, each object inserted into the
tree is a concatenation of two codewords of size-$\left(\left| G\right| -1\right)$\ restrictions.

After the codewords are created, a second phase of the algorithm deletes redundant
$\operatorname*{AND}_{k}$, $\operatorname*{OR}_{k}$, and $\left(\urcorner\right)\operatorname*{XOR}_{k}$\ vertices. \ This phase looks
for vertices $u$\ and $v$ with $L\left(  u\right)  $\ and $L\left(  v\right)
$ incomparable, which, as a consequence of Theorem \ref{unique}, can only have arisen by
$\operatorname*{AND}_{k}$, $\operatorname*{OR}_{k}$, or $\left(
\urcorner\right)  \operatorname*{XOR}_{k}$.
\ Both phases effectively perform an\ $O\left(  \log
N\right)  $-time operation for all subsets of subsets of $V_{f}$, so
the complexity is $O(N^{1.585}\log N)$.
\vspace{-0.3cm} \section{\label{quantumsec}Quantum Query Complexity}
\vspace{-0.4cm} \noindent The quantum query complexity\ of a Boolean function $f$ is the minimum number
of oracle queries needed by a quantum computer to evaluate $f$. \ Here we are
concerned only with the bounded-error query complexity $Q_{2}\left(
f\right)  $ (def\hspace{0pt}ined in \cite{bbcmw}), since approximating unitary matrices with f\hspace{0pt}inite precision introduces 
bounded error into any quantum algorithm. 
A quantum query algorithm $\Gamma$ proceeds by an alternating
sequence of $T+1$ unitary transformations and $T$ query
transformations:
$U_{0}\rightarrow Q_{1}\rightarrow U_{1}\rightarrow\cdots\rightarrow
Q_{T}\rightarrow U_{T}.$
Then $Q_{2}(f)$ is the minimum
of $T$ over all $\Gamma$ that compute $f$ with bounded error.

There are several open problems that an ef\hspace{0pt}f\hspace{0pt}icient algorithm to compute $Q_{2}(f)$ might help to investigate
\cite{ambainis,bbcmw,bw}.  \ Unfortunately, we do not know of such an algorithm. \ Here we show that, if
we limit the number of qubits, we can obtain a subexponential-time 
approximation algorithm via careful
exhaustive search.
\vspace{-0.2cm} \subsection{\label{overviewsec}Overview of Result}
\vspace{-0.4cm} \noindent For what follows, it will be convenient to extend the quantum oracle model to
allow intermediate observations. \ With an unlimited workspace,
this cannot change the number of queries needed
\cite{bv}.  In the space-bounded setting, however, it might make a
difference.

We def\hspace{0pt}ine a \textit{composite algorithm} $\Gamma^{\prime}$\ to be an alternating sequence
$ \Gamma_{1}\rightarrow D_{1}\rightarrow\cdots\rightarrow \Gamma_{t}\rightarrow D_{t}. $
Each $\Gamma_{i}$\ is a quantum query algorithm that uses $T_{i}$\ queries and at most $m$ qubits of memory for some $m\geq\log_{2}%
n+2$. \ When $\Gamma_{i}$ terminates a basis state $\left|  \psi_{i}\right\rangle$ is observed. 
Each $D_{i}$\ is a \textit{decision point}, which takes as input the sequence
$\left|  \psi_{1}\right\rangle ,\ldots,\left|  \psi_{t}\right\rangle $, and as
output decides whether to (1) halt and return $f=0$, (2) halt and return  $f=1$, or
(3) continue to $\Gamma_{t+1}$.
(The f\hspace{0pt}inal decision point, $D_{t}$, must select between (1) and (2).) \ There
are no computational restrictions placed on the decision points. \ However, a
decision point cannot modify the quantum algorithms that come later in
the sequence; it can only decide whether to continue with the sequence.
For a particular input, let $p_{k}$ be the probability, over all runs of $\Gamma^{\prime}$, that quantum algorithm $\Gamma_{k}$
is invoked.  Then $\Gamma^{\prime}$ uses a total number of queries
$ {\sum_{k=1}^{t}} p_{k} T_{k}. $

We def\hspace{0pt}ine the space-bounded quantum query complexity
$SQ_{2,m}(f)$ to be the minimum number of queries used by any composite
algorithm that computes $f$ with error probability at most $1/3$ and that is restricted to $m$ qubits.
\ We give an approximation algorithm for
$SQ_{2,m}(f)$\ taking time $2^{O\left( 4^{m}mn\right) }$, which when
$m=O\left(  \log n\right)  $ is $O( N^{\operatorname*{polylog}%
(N)}) $. \ The approximation ratio is $\sqrt{22}/3+\epsilon$ for any $\epsilon>0$. \ The dif\hspace{0pt}f\hspace{0pt}iculty in proving the result is as follows.

A unitary transformation is represented by a continuous-valued
matrix, which might suggest that the quantum model of computation is analog
rather than digital. \ But Bernstein and Vazirani \cite{bv} showed that, for a
quantum computation taking $T$ steps, the matrix entries need to be accurate only to
within $O(\log T)$ bits of precision in the bounded-error model. \ However,
when we try represent unitary transformations on a computer with f\hspace{0pt}inite
precision, a new problem arises. \ On the one hand, if we allow only matrices
that are exactly unitary, we may not be able to approximate every unitary
matrix. \ So we also need to admit matrices that are \textit{almost} unitary.
\ For example, we might admit a matrix if the norm of each row is suf\hspace{0pt}f\hspace{0pt}iciently
close to $1$, and if the inner product of each pair of distinct rows is
suf\hspace{0pt}f\hspace{0pt}iciently close to $0$. \ But how do we know that every such matrix is
close to some actual unitary matrix? \ If it is not, then the transformation it
represents cannot even approximately be realized by a quantum computer.

We resolve this issue as follows. \ F\hspace{0pt}irst, we show that every almost-unitary
matrix is close to some unitary matrix in a standard metric. \ Second, we
show that every unitary matrix is close to some almost-unitary matrix representable with limited precision. 
\ Third, we upper-bound the precision that suff\hspace{0pt}ices for a
quantum algorithm, given a f\hspace{0pt}ixed accuracy that the algorithm
needs to attain.

An alternative approach to approximating $SQ_{2,m}(f)$\ would be to represent
each unitary matrix as a product of elementary gates. \ Kitaev \cite{kitaev}%
\ and independently Solovay \cite{solovay}\ showed that a $2^{m}\times2^{m}%
$\ unitary matrix can be represented with arbitrary accuracy $\delta>0$\ by a
product of $2^{O(m)\operatorname*{polylog}(1/\delta)}$\ unitary gates. \ But
this yields a $2^{2^{ O(m)\operatorname*{polylog}\left(
mn\right) }  }$\ algorithm, which is slower than ours. \ Perhaps the construction or its
analysis can be improved; in any case, though, this approach is not as
natural for the setting of query complexity.
\vspace{-0.2cm} \subsection{\label{almostsec}Almost-Unitary Matrices}
\vspace{-0.2cm} \noindent Let $u\bullet v$ denote the conjugate inner product of $u$ and $v$.
\ The distance $\left|A-B\right|$ between matrices $A=\left(  a_{ij}\right)  $\ and $B=\left(
b_{ij}\right)  $\ in the $L_{\max}$\ norm is def\hspace{0pt}ined to be $\max_{i,j}\left|
a_{ij}-b_{ij}\right|  $.
\begin{definition}
A matrix $A$ is $q$-{\normalfont almost-unitary} if $|I-AA^{\dagger}|<q$.
\end{definition}
In the following lemma, we start an
almost-unitary matrix $A$ and construct an actual unitary matrix $U$ that is close to
$A$ in the $L_{\max}$ norm.

\begin{lemma}
\label{u1}Let $A$ be a $q$-almost-unitary $s\times s$
matrix, with $s\geq 2$ and $q \leq 1/4s$. Then there exists a unitary matrix $U$
such that $\left|A-U\right| < 4.91q\sqrt{s}$.
\end{lemma}
\begin{proof}
We f\hspace{0pt}irst normalize each row $A_{i}$ so that  $A_{i}\bullet A_{i}=1$. For
each entry $a_{ij}$,
$ |a_{ij}/(A_{i}\bullet A_{i})-a_{ij}| = |a_{ij}||1-(A_{i}\bullet A_{i})|/|A_{i}\bullet A_{i}|
< q\left(1+q\right)/\left(1-q\right). $
We next form a unitary matrix $B$ from $A$ by using the Classical Gram-Schmidt
(CGS) orthogonalization procedure (see \cite{hk} for details). \ The idea is to project $A_{2}$ to make it orthogonal to $A_{1}$,
then project $A_{3}$ to make it orthogonal to both $A_{1}$ and $A_{2}$, and so on. \
Initially we set $B_{1}\leftarrow A_{1}$. \ Then for each $2\leq i\leq s$,
we set
$ B_{i}\leftarrow A_{i}-{\sum_{j=1}^{i-1}(A_{i}\bullet B_{j})B_{j}}$.
Therefore $A_{i}\bullet B_{k} = 
(A_{i}\bullet A_{k}) - {\textstyle\sum_{j=1}^{k-1}(A_{i}\bullet B_{j})(A_{k}\bullet B_{j})}. $

We need to show that the discrepancy between $A$ and $B$ does not increase too
drastically as the recursion proceeds.  Let $\sigma_{k}=\max_{i} A_{i}\bullet B_{k}$.
\ By hypothesis, $\sigma_{1} < q$.  Then
$\sigma_{k}\leq \sigma_{1}+{\sum_{j=1}^{k-1} \sigma_{j}^{2}}$.
Assume that $\sigma_{k}<q+4q^{2}s$ for all $k \leq K$.  By induction,
$\sigma_{K+1}< q+K\left( q+4q^{2}s\right)^{2} \leq q+4q^{2}s$
since $q \leq 1/4s$ and $K \leq s$.  So for all $k$, $\sigma_{k}<q+4q^{2}s$.

Let $\phi = \left|A-B\right|$.  By the def\hspace{0pt}inition of $B$,
$ \phi \leq \sigma_{1} \left| w_{1}\right| + \cdots + \sigma_{s} \left| w_{s}\right| $
where $w$ is a column of $B$.  Since $\left| w_{1}\right| ^{2} + \cdots + \left| w_{s}\right| ^{2}=1$, $\phi$ is maximized when 
$w_{i} = \sigma_{i} \sqrt{s} / \left( \sigma_{1} + \cdots + \sigma_{s} \right)$, or 
$\phi \leq \sigma_{1}^{2} + \cdots + \sigma_{s}^{2} \sqrt{s} / (\sigma_{1}+\cdots +\sigma_{s}) \leq 
\left( q+4q^{2}s \right)^{2} \sqrt{s}/q$.

Adding $q\left(1+q\right)/\left(1-q\right)$ from
normalization yields a quantity less than \linebreak $\left(4+9\sqrt{2}/14\right) q \sqrt{s} \approx 4.91q\sqrt{s}$.  
This can be seen by working out the arithmetic for the worst case of
$s=2$, $q=1/4s$.
\end{proof}

The next lemma, which is similar to Lemma 6.1.3 of \cite{bv}, is a sort of converse to Lemma \ref{u1}: we start with an
arbitrary unitary matrix, and show that truncating its entries to a precision $\delta>0$ produces an
almost-unitary matrix.

\begin{lemma}
\label{u2}
Let $U$ and $V$ be $s\times s$ matrices with $s\geq 2$ and $\left|U-V\right| < \delta$.
If $U$ is unitary, then $V$ is $\left(  2\delta\sqrt{s}+\delta^{2}s\right)$-almost-unitary.
\end{lemma}
\begin{proof}
\nobreak F\hspace{0pt}irst, $ U_{i}\bullet U_{i} = \sum_{k=1}^{s}\left|u_{k}+\gamma_{k}\right|^{2} = 
1+\sum_{k=1}^{s} \left( u_{k} \gamma_{k}^{\ast}+u_{k}^{\ast}\gamma_{k} + \gamma_{k}\gamma_{k}^{\ast} \right)$
where the $u_{k}$'s are entries of $U$ and the $\gamma_{k}$'s are error terms satisfying
$|\gamma_{k}|<\delta$. \ So by the Cauchy-Schwarz inequality,
$U_{i}\bullet U_{i}$ differs from $1$ by at most $2\delta\sqrt{s}+\delta^{2}s$.  Second, for 
$i\neq j$, $ U_{i}\bullet U_{j}=\sum_{k=1}^{s}(u_{k}+\gamma_{k})(u_{k}+\eta_{k})^{\ast}$ 
where the $\gamma_{k}$'s and $\eta_{k} $'s are error terms, and the argument proceeds analogously.
\end{proof}

\vspace{-0.2cm} \subsection{\label{searchsec}Searching for Quantum Algorithms}
\vspace{-0.2cm} \noindent In this section we use the results on almost-unitary matrices to construct an
algorithm. \ F\hspace{0pt}irst we need a lemma about error buildup in quantum algorithms,
which is similar to Corollary 3.4.4\ of \cite{bv} (though the proof technique is different).

\begin{lemma}
\label{u3}Let $U_{1},\ldots U_{T}$ be $s\times s$ unitary matrices,
 $\widehat{U_{1}},\ldots\widehat{U_{T}}$\ be $s\times s$ arbitrary matrices,
and $v$ be an $s\times1$\ vector with $\left\|
v\right\|  _{2}=1$. \ Suppose that, for all $i$,
$\left|\widehat{U_{i}} - U_{i}\right|<1/cs$,
where $c> T/2$. \ Then $\widehat{U_{1}}\cdots
\widehat{U_{T}}v$\ differs from $U_{1}
\cdots U_{T}v$ by at most $2T/\left[  \sqrt{s}\left(
2c-T\right)  \right]  $\ in the $L_{2}$ norm.
\end{lemma}
\begin{proof}
\nobreak For each $i$, let $E_{i}=\widehat{U_{i}}-U_{i}$.  By hypothesis, every entry of $E_{i}$ has magnitude at most $1/cs$; 
thus, each row or column $w$ of $E_{i}$\ has $\left\|  w\right\| _{2}\leq 1/\left(c\sqrt{s}\right)$. \ Then 
$ \widehat{U_{1}}\cdots\widehat{U_{T}}v=\left(
U_{1}+E_{1}\right)  \cdots\left(  U_{T}+E_{T}\right)
v. $
The right-hand side, when expanded, has $2^{T}$ terms. \ Any term containing
$k$ matrices $E_{i}$\ has $L_{2}$\ norm at most $s^{-1/2} c^{-k}$,
and can therefore add at most $c^{-k}/\sqrt{s}$ to the discrepancy with $U_{1}\cdots U_{T}v$.
\ So the total discrepancy is at most
$ s^{-1/2} {\sum_{k=1}^{T}} \tbinom{T}{i}\left(  1/c\right)  ^{k}< s^{-1/2} \left(  e^{T/c}-1\right). $
Since $d\ln t/dt$ evaluated at $t=2c$\ is $1/{2c}$ and since $\ln t$ is concave, 
$ \ln(2c+T)-\ln(2c-T)\geq2T/2c  =T/c $
when $T<2c$. \ Therefore
$ e^{T/c} \leq (2c+T)/(2c-T) $
and the discrepancy is at most $2T/\left[  \sqrt{s}\left(  2c-T\right)
\right]$ in the $L_{2}$\ norm.
\end{proof}

Applying Lemmas \ref{u1}, \ref{u2}, and \ref{u3}, we now prove the main theorem.

\begin{theorem}
There exists an approximation algorithm for $SQ_{2,m}(f)$\ taking time
$2^{O\left(  4^{m}mn\right) }$, with approximation ratio $\sqrt{22}/3+\epsilon$.
\end{theorem}
\begin{proof}
Given $f$, we want, subject to the following
two constraints, to f\hspace{0pt}ind an algorithm $\Gamma$ that
approximates $f$ with a minimum number of queries. \ F\hspace{0pt}irst, $\Gamma$ uses at most $m$ qubits,
meaning that $s=2^{m}$ and the relevant matrices are $2^{m}\times
2^{m}$. \ Second, the correctness probability of $\Gamma$ is known to a constant
accuracy $\pm\varepsilon$.
Certainly the number $T$ of queries never needs to be more than $n$, for, although each quantum algorithm is space-bounded,
the {\it composite} algorithm need not be. \ Let
$\lambda$\ be the $L_{\max}$ error we can tolerate in the matrices, and let
$\Delta$ be the resultant $L_{2}$ error in the f\hspace{0pt}inal states. \
Setting $c=1/\left( \lambda 2^{m}\right)$, by Lemma \ref{u3} we have
$ \Delta \leq 2n/\left[  2^{m/2}\left(  2^{1-m}/\lambda-n\right)\right].$
From the Cauchy-Schwarz inequality, one can show that $\varepsilon \leq 2\Delta$.  Then solving for $1/\lambda$,
$ 1/\lambda \leq 2^{m/2}n\left( 2/\varepsilon+1 \right) $
which, since $\varepsilon$ is constant, is $O\left(  2^{m/2}n\right)$.  Solving for $c$, we can verify that $c> T/2$, 
as required by Lemma \ref{u3}. 
If we generate almost-unitary matrices, they need to be within
$\lambda$ of actual unitary matrices. \ By Lemma \ref{u1} we can use
 $\lambda/\left( 4.91\sqrt{s} \right)$-almost-unitary matrices. 
F\hspace{0pt}inally we need to ensure that we approximate every
unitary matrix. \ Let $\delta$ be the needed precision. \ Invoking Lemma \ref
{u2}, we set
$ \lambda/\left( 4.91 \sqrt{s}\right) \geq2\delta\sqrt{s}+\delta^{2}s $
and obtain that
$ \delta\leq\max\left[ \lambda/\left( 9.82s \right)  ,\lambda^{1/2}
/\left( 2.22 s^{3/4}\right)\right] $
is suff\hspace{0pt}icient.

Therefore the number of bits of precision needed per entry, $\log\left(
1/\delta\right)  $, is $O(m)$. \ We thus need only $O(4^{m}
mn)$\ bits to specify $\Gamma$, and can search through all possible $\Gamma$ in time 
$2^{O(4^{m} mn)}$. \ The amount of time
needed to evaluate a composite algorithm $\Gamma^{\prime}$ is polynomial in $m$ and
$n$, and is absorbed into the exponent. \
The approximation algorithm is this: f\hspace{0pt}irst let $\varepsilon>0$ be
a constant at most $0.0268$, and let $ \omega=\frac{22} {9}+\frac{4}{3}\varepsilon-8\varepsilon^{2}.$
\ Then f\hspace{0pt}ind the smallest $T$ such that the maximum probability of correctness
over all $T$-query algorithms $\Gamma^{\prime}$\ is at least $2/3-\varepsilon$ (subject to $\pm\varepsilon
$ uncertainty), and\ return $T\sqrt{\omega} $. \ The algorithm achieves an approximation ratio of $\sqrt{\omega
} $, for the following reason. \ F\hspace{0pt}irst, $T\leq SQ_{2,m} (f)$.
\ Second, $\omega T\geq SQ_{2,m}(f)$, since by repeating the optimal algorithm
$\Gamma^{\ast}$\ until it returns the same answer twice (which takes either two or three repetitions), the correctness probability can be boosted above $2/3$. \ 
 F\hspace{0pt}inally, a simple calculation reveals that
$\Gamma^{\ast}$\ returns the same answer twice after expected number of
invocations $\omega$.
\end{proof}
\vspace{-0.3cm} \section{Acknowledgments}
\vspace{-0.4cm} \noindent I thank Umesh Vazirani for advice and encouragement, Rob Pike and Lorenz Huelsbergen for 
sponsoring the Bell Labs
internship during which this work was done and for helpful discussions, Andris Ambainis and an anonymous reviewer for comments and corrections,
Wim van Dam for a simplif\hspace{0pt}ication in Section 7, and
Peter Bro Miltersen for correspondence.


\begin{thebibliography}{99}
\vspace{-0.2cm} \bibitem{aaronson}S. Aaronson,\ Boolean Function Wizard 1.0 (software
library),\ http://www.cs.berkeley.edu/\symbol{126}aaronson/bfw, 2000.

\bibitem{ambainis}A. Ambainis,\ \textit{Quantum lower bounds by quantum
arguments},\ in Proceedings of the Thirty-Second Annual ACM Symposium on
Theory of Computing, ACM, Portland, OR, 2000, pp. 636--643.

\bibitem{bbcmw}R. Beals, H. Buhrman, R. Cleve, M. Mosca, and R. de
Wolf,\ \textit{Quantum lower bounds by polynomials},\ in Proc. 39'th IEEE
Symp. on Foundations of Comp. Sci., 1998, pp. 352--361.

\bibitem{bv}E. Bernstein and U. Vazirani,\ \textit{Quantum complexity
theory},\ SIAM J. Comput., 26:5(1997), pp. 1411--1473.

\bibitem{bw}H. Buhrman and R. de Wolf,\ \textit{Complexity measures and decision
tree complexity: a survey},\ to appear in Theoretical Comp. Sci.

\bibitem{czort}S. L. A. Czort,\ \textit{The complexity of minimizing disjunctive normal form formulas},\
Master's Thesis, University of Aarhus, 1999.

\bibitem{glr}D. Guijarro, V. Lav\'{i}n, and V. Raghavan, \textit{Exact learning when irrelevant variables abound},
Information Proc. Lett., 70(1999), pp. 233--239.

\bibitem{hjlt}T. Hancock, T. Jiang, M. Li, and J. Tromp, \textit{Lower bounds on learning decision lists and trees},
Information and Computation, 126(1996), pp. 114--122.

\bibitem{hk}K. Hoffman and R. Kunze,\ \textit{Linear Algebra},\ Prentice Hall, 1971.

\bibitem{hr}L. Hyaf\hspace{0pt}il and R. L. Rivest, \textit{Constructing optimal binary decision trees is NP-complete},
Information Proc. Lett., 5(1976), pp. 15--17.

\bibitem{kc}V. Kabanets and J-Y Cai,\ \textit{Circuit minimization
problem},\ in Proceedings of the Thirty-Second Annual ACM Symposium on Theory
of Computing, ACM, Portland, OR, 2000, pp. 73--79.

\bibitem{kitaev}A. Yu. Kitaev,\ \textit{Quantum computations: algorithms and
error correction},\ Russian Math. Surveys, 52:6(1997), pp. 1191--1249.

\bibitem{nisan}N. Nisan,\ \textit{CREW PRAMs and decision trees},\ SIAM J.
Comput., 20:6(1991), pp. 999--1007.

\bibitem{ns}N. Nisan and M. Szegedy,\ \textit{On the degree of Boolean
functions as real polynomials},\ Comput. Complexity, 4:4(1994), pp. 301--313.
\ Earlier version in STOC'92.

\bibitem{rr}A. A. Razborov and S. Rudich,\ \textit{Natural proofs},\ J.
Comput. System Sci., 55(1997), pp. 24--35.

\bibitem{rubinstein}D. Rubinstein,\ \textit{Sensitivity vs. block sensitivity
of Boolean functions},\ Combinatorica, 15:2(1995), pp. 297-299.

\bibitem{sw}M. Saks and A. Wigderson,\ \textit{Probabilistic Boolean decision
trees and the complexity of evaluating game trees},\ in Proceedings of the
Twenty-Seventh IEEE Symposium on Foundations of Computer Science, IEEE
Computer Society, Ontario, Canada, 1986, pp. 29--38.

\bibitem{solovay}R. Solovay,\ \textit{Lie groups and quantum circuits},\ talk
at workshop on Mathematics of Quantum Computation, Mathematical Sciences
Research Institute, Spring 2000.

\bibitem{trakhtenbrot}B. A. Trakhtenbrot,\ \textit{A survey of Russian
approaches to perebor (brute-force search) algorithms},\ Annals of the History
of Computing, 6:4(1984), pp. 384--400.

\end{thebibliography}
\end{document}